\documentclass[a4paper,fleqn,usenatbib]{mnras}
\usepackage{amssymb,amsmath,graphicx,psfrag,mathtools}
\usepackage[T1]{fontenc} 
\usepackage{ae,aecompl} 
\usepackage{url}
\usepackage{revsymb}
\hypersetup{draft} 
\title[Aperiodicity in Repeating FRB]{The Absence of Periodicity in Repeating FRB
}
\author[J. I. Katz]{
J. I. Katz,$^{1}$\thanks{E-mail katz@wuphys.wustl.edu} 
\\
$^{1}$Department of Physics and McDonnell Center for the Space Sciences,
Washington University, St. Louis, Mo. 63130 USA 
}
\date{Accepted XXX.  Received YYY; in original form ZZZ} 
\pubyear{2022} 
\date{\today}
\begin{document} 
\label{firstpage} 
\pagerange{\pageref{firstpage}--\pageref{lastpage}} 
\maketitle 
\begin{abstract}
Popular Fast Radio Burst models involve rotating magnetized neutron stars,
	yet no rotational periodicities have been found.  {Small
	datasets exclude exact periodicity in FRB 121102.  Recent}
	observations of over 1500 bursts from each of FRB 121102 and FRB
	20201124A {have also not found periodicity}.  Periodograms of
	events with cosine-distributed random offsets as large as $\pm 0.6
	P$ from a strict period $P$ would {still} reveal the underlying
	periodicity.  {The sensitivity of periodograms of long data series,
	such as bursts observed on multiple days, to slow frequency drifts
	is mitigated by considering individual observing sessions, and
	results are shown for FRB 121102.} Models of repeating FRB without
	intrinsic periodicity are considered, as are models of apparently
	non-repeating FRB.
\end{abstract}
\begin{keywords} 
radio continuum, transients: fast radio bursts, accretion, accretion discs,
stars: black holes, stars: magnetars
\end{keywords} 
\section{Introduction}
The sources of Fast Radio Bursts (FRB) remain mysterious.  Strongly
magnetized neutron stars (``magnetars'') have long been proposed because
their great magnetostatic energy is believed to be released in Soft Gamma
Repeater (SGR) outbursts \citep{K82}, and because neutron stars have the
short characteristic time scales (manifested in the sub-ms rise times of SGR
and in pulsar pulse widths and substructure) required to explain FRB.  As
radio pulsars, magnetic neutron stars radiate coherently with
extraordinarily high brightness temperatures, another property of FRB.
Although the giant outburst of the Galactic SGR 1806$-$20 did not produce a
FRB \citep{TKP16}, setting an upper bound on its isotropic-equivalent FRB
energy about 11 orders of magnitude lower than that of FRB at redshifts
$z \sim 1$, related objects, perhaps with different values of their
parameters, are popular models of FRB sources \citep{P18}.

The magnetically mediated or powered emission of any rotating magnetized
object must be periodic at its rotational frequency, unless the magnetic
field is (implausibly) accurately dipolar and accurately aligned with the
rotational axis.  Radio pulsars are the classic example, and the pulses of
RRAT (radio pulsars most of whose pulses are nulled) are separated by
integer multiples of their underlying (rotational) periods.  {This
applies even if the radiation is produced by a collimated relativistic beam
far from the neutron star \citep{M19} because the direction of the beam and
its radiation are tied to the orientation of the rotating neutron star.}
Even the thermal emission of Anomalous X-ray Pulsars (AXP; the quiescent
counterparts of SGR) is periodic with their rotational period, as is the
emission of accreting binary neutron stars.

Two types of periodicity may be considered: bursts separated by integer
multiples of a stable underlying period (as in radio PSR and RRAT) and
near-periodic modulation of activity.  The latter describes a process that
has irregular, perhaps random, scatter about the underlying stable period;
an example is observed Solar activity that is modulated at the underlying
more stable Solar rotation period.  Either would produce a narrow spike in a
periodogram and in the distribution of burst intervals.

Despite the expectation of rotational periodicity in the activity of
repeating FRB, no such periodicity has been found.  Long period modulation
of the activity of two FRB has been observed (16.35 d of FRB 180916
\citep{CHIME20} and 160 d of FRB 121102 \citep{R20}), but these periods are
too long to be plausibly identified as neutron star rotational periods.
Many repetitions of repeating FRB have recently been reported, including
1652 bursts of FRB 121102 \citep{L21a} and 1863 bursts of FRB 20201124A
\citep{X21}, in both cases without a spike in periodograms {extending
over the approximately two months of observation} or other evidence of
periodicity.  Are these results consistent with an underlying periodicity,
as required in a magnetic neutron star model because neutron stars rotate,
or do they point to entirely different models?

{Orbital motion and spindown may interfere with a search for periodic
behavior in a rotating neutron star model by introducing large phase
offsets.  Popular neutron star models of FRB assume them to resemble Soft
Gamma Repeaters (SGR; ``magnetars''), none of which are binary.  For this
reason and because of the difficulty of searching the large phase space of
possible binary orbits and time-dependence of the resulting phase shifts,
this paper assumes a single object.}

{I first describe an unsuccessful search for exact periodicity in a
small sample of bursts so closely spaced in time that period changes are
unlikely to be significant.  Next, I consider the effects of frequency
drifts, such as would be produced by neutron star spin-down (or spin-up).
These effects are mitigated by analyzing each comparatively brief
observing session independently; no periodicity is found.  Then I calculate}
the periodogram of a simple model in which an underlying clock is stable,
but bursts are randomly distributed in phase about its period.  The
periodogram reveals the presence of the clock even when the random phase
deviations, distributed by a cosine probability function, may equal or
slightly exceed $\pm$ a half cycle.  The absence of {significant peaks}
in the periodograms {calculated here using the data of \citet{L21a}
(unfortunately, \citet{X21} have not yet published or made available their
data)} is evidence that there is no such clock in FRB 121102, and supports
arguments \citep{K20} against magnetic neutron star models.  Alternative
models of repeating and non-repeating FRB are considered in
Sec.~\ref{alternatives}.
\section{Exactly Periodic Bursts}
\label{exact}
{\citet{G18} observed five bursts (11B--F) of FRB 121102 within
0.001072338 d (93 s).  For a neutron star spinning down by emission of
magnetic dipole radiation (a similar rate is expected for an aligned rotor)
the rate of change of spin frequency at age $A$
\begin{equation}
	\label{omegadot}
	\begin{split}
		{\dot \omega} &= - {\omega \over 2 A}\\
		\left|{\dot \omega}\right| &\lesssim 2 \times 10^{-5}\
		\text{s}^{-2},
	\end{split}
\end{equation}
where $\omega < 2\pi\times 10^3\,$s (period $P > 1\,$ms) and $A > 5\,$y
have been assumed (this bound on $A$ is applicable to the data of
\citet{G18}) and it is assumed the present spin period is much longer than
the spin period at birth (if this is not true, then a yet stricter bound
results).  This result is more generally applicable if $A$ is taken as the
spin-down age, that might greatly exceed the actual age.

The resulting phase drift, measured from the middle of the observing
interval of length $T = 93\,$s, is
	\begin{equation}
		\label{Deltaphi}
		\Delta \phi = {1 \over 8} {\dot \omega} T^2 \lesssim 0.02\
		\text{radian}.
	\end{equation}
Even for the fastest and youngest possible rotating neutron star, spindown
cannot interfere with detecting exact periodicity in this 93 s interval,
were it present.

Combining Eqs.~\ref{omegadot}, \ref{Deltaphi} yields the maximum useful
data span if $\dot\omega$ is constant:
\begin{equation}
	T < \sqrt{16 \Delta\phi A \over \omega} \approx 1.6 \times 10^3
	\sqrt{{A \over 10\, \text{y}}{2 \pi \times 10^3\,\text{s}^{-1}
	\over \omega}}\,\text{s},
\end{equation}
where $\Delta \phi \approx \pi$ is the maximum phase drift consistent with
finding a significant peak in the periodogram.  It is possible to
generalize the periodogram and increase its power by adding an $\dot\omega$
term like that in Eq.~\ref{Deltaphi} to the phase $\phi$, but that would
increase the dimensionality of the parameter space that must be searched.

An exercise in Diophantine arithmetic shows that the four independent
intervals between bursts 11B, 11C, 11D, 11E and 11F are not consistent with
integer multiples of a constant underlying period.  Alternatively, the
space of possible rotational frequencies from 20/d to $2 \times 10^8$/d
($231\,\mu$Hz to 2.31 kHz) is evenly sampled, corresponding to periods of
0.432 ms to 1.2 hours.  A figure of merit may be defined:
\begin{equation}
	\label{FOMeq}
	\text{FOM}(P) = \sum_{i=\text{C,D,E,F}}
	\left({(t_i - t_\text{B})-\text{NINT}[(t_i - t_\text{B})/P]P
	\over P}\right)^2,
\end{equation}
where NINT is the nearest integer function, the quantity in large
parentheses is the deviation in units of $P$ of $t_i - t_\text{B}$
($t_\text{B}$ functions as a reference time) from an integer multiple of
$P$ and FOM measures the deviation of the four independent intervals from
exact periodicity.

The resulting distribution of FOM is shown in Fig.~\ref{FOMfig}.  Some
values of $P$ will, entirely fortuitously, provide a good fit to the
observed intervals even if there is no underlying periodicity.  Because the
frequency resolution is greater than the maximum physically possible
frequency (the limiting rotation rate of a neutron star), and orders of
magnitude greater than likely frequencies (SGR rotation rates), a true
periodicity would appear as multiple occurrences of very small
r.m.s.~deviations from the smooth curve expected for uncorrelated pulses.
No such excess is observed.
\begin{figure}
	\centering
	\includegraphics[width=0.99\columnwidth]{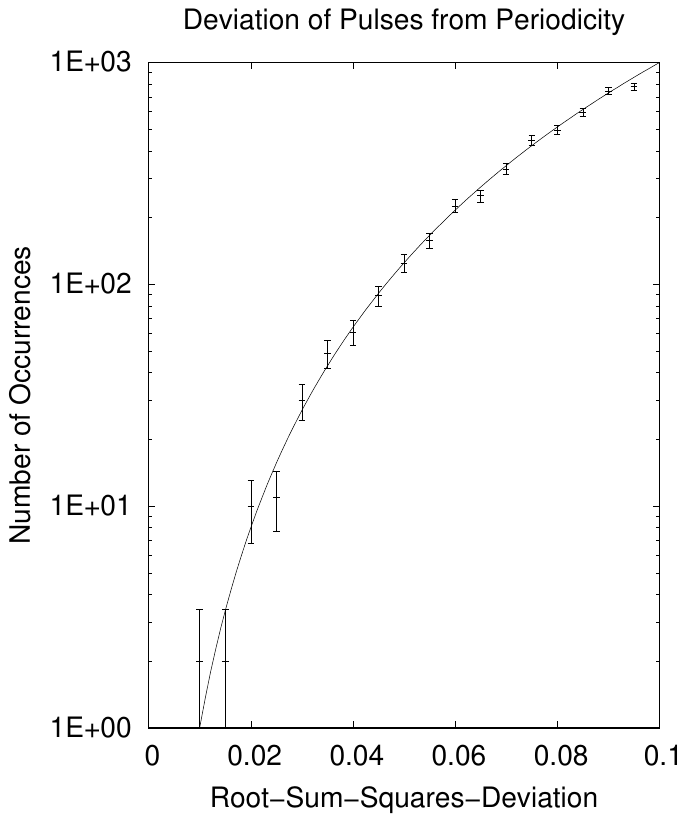}
	\caption{\label{FOMfig} Distribution of r.m.s.~deviations of
	times of bursts 11B--F of \citet{G18} from integer multiples of
	exact periods, for $10^7$ periods evenly spaced in frequency from
	231 $\mu$Hz to 2.31 kHz (periods 0.432 ms to 1.2 hr).  Error bars
	are $1\sigma$.  A smooth curve, as observed, is expected for
	uncorrelated aperiodic bursts.  Exact periodicity would appear as a
	period showing zero (except for measurement and roundoff errors)
	deviation; none is found.}
\end{figure}
}
\section{The Problem of Frequency Drift}
{Unfortunately, calculating periodograms of entire datasets extending
over months, as was done by \citet{L21a,X21}, may not reveal a periodicity
because even tiny period derivatives can dephase bursts months apart.  This
is a particular problem for fast (ms) rotation periods, but would be less so
for the multi-second periods of known SGR/AXP.  For example,
Eq.~\ref{Deltaphi} shows that for $T \approx 2\,$months, as in the datasets
of \cite{L21a,X21}, $|\dot\omega| = 2 \times 10^{-12}\,$s$^{-2}$ is
sufficient to dephase by $2\pi$ radians.  For comparison, most SGR/AXP have
$|\dot\omega| \gtrsim 10^{-11}\,$s$^{-2}$.  If repeating FRB are made by
SGR-like objects, periodograms of months-long datasets, as computed by
\citet{L21a,X21}, may not reveal their periodicity.

This problem may be mitigated by computing the periodograms of individual
observing sessions.  The long datasets consist of many shorter observing
sessions as the sources pass through the field of view of FAST, a transit
instrument.  For example, most of the observing sessions of \citet{L21a}
are one hour long, although a few are as long as five hours.  These
may be considered individually, and their average may reveal low-amplitude
modulation not apparent in data from individual sessions (although
frequency drift may move the signal to different periodogram bins in
different sessions).

For $T = 1\,$hour, dephasing by $\Delta \phi = \pi$ radians only occurs if 
\begin{equation}
	\left|{\dot\omega}\right| \ge {8 \Delta\phi \over T^2} \approx
	2 \times 10^{-6}\,\text{s}^{-2}.
\end{equation}
Such large values of $|{\dot\omega}|$ are possible if $\omega = 2 \pi \times
10^3\,$s$^{-1}$ and $A = 10\,$y, but only at these extremes of both
parameter ranges.

\citet{L21a} detected bursts in 39 distinct observing sessions, spread over 
about two months.  The number of bursts in a single session ranged from one
to 122 (an additional eight sessions detected no bursts).  Of those 39
sessions, the 17 with at least 50 bursts (to ensure good statistics in the
periodograms) were analyzed; these comprise 78\% of the total 1652 bursts.
The 17 individual periodograms, evaluated at the $3.6 \times 10^6$ evenly
spaced frequencies from 1/h to $10^3$/s, were then averaged.

There was no evidence of a periodicity in any of the individual periodograms
or in their average.  Because it is difficult to display graphically the
$3.6 \times 10^6$ elements of an individual periodogram, the distribution of
averages of the amplitudes of the 17 single-session periodograms is shown in
Fig.~\ref{121102mean}.  Examination of the highest averages also shows no
evidence for periodicity (which might occur at slightly different periods in
the individual sessions, and be smoothed in the average): The greatest
amplitude is only 3\% greater than the second-highest, and they are at very
different periods, not indicating a smoothed or broadened periodicity.

\begin{figure}
	\centering
	\includegraphics[width=0.99\columnwidth]{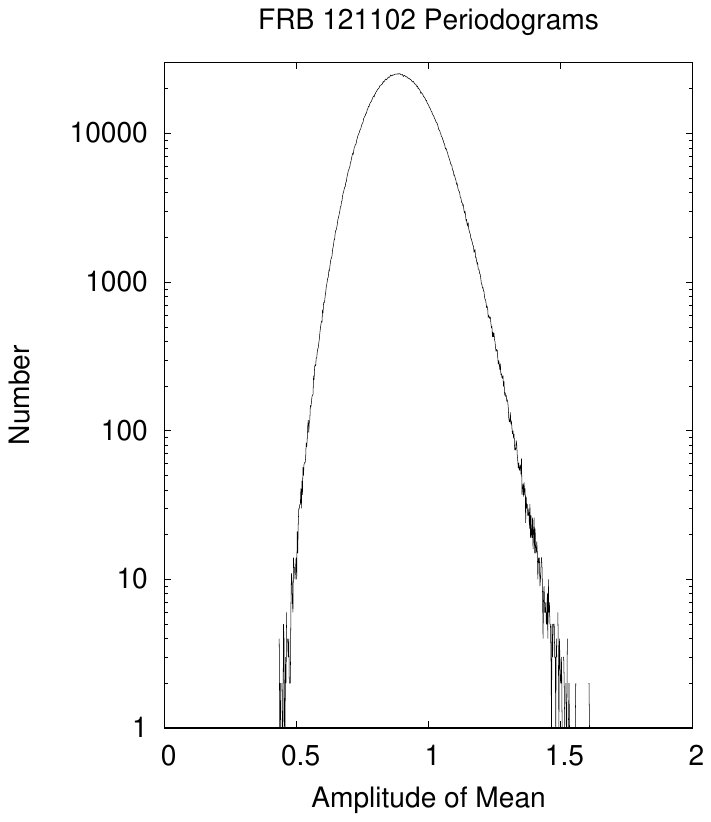}
	\caption{\label{121102mean} Distribution of the the $3.6 \times
	10^6$ averages of the amplitudes of 17 single-session (sessions with
	$\ge 50$ bursts) periodograms of the bursts of FRB 121102 for
	periods from 1 ms to 1 hour; data from \citet{L21a}.  The amplitude
	normalization is arbitrary.  The distribution is close to the
	Gaussian expected for ``shot noise'' burst times, somewhat broadened
	and skewed by the slow variations in activity of FRB 121102.  The
	largest average amplitude in the data is 1.706; there is no evidence
	of periodicity.}
\end{figure}

The 17 individual single-session periodograms, that would not be expected
to be significantly affected by period drift, also show no evidence of
periodicity.  None of the highest amplitudes exceed the second-highest by
more than 11\%, and in each periodogram the highest amplitudes are at very
different periods, unlike a broadened peak.  The top ten are shown for each
session and for the average in the Appendix.

The amplitudes of the largest values of the average periodogram are smaller
than those of the individual session periodograms because extremes are
reduced by averaging with (smaller) values of the other periodograms at the
same frequencies.  The highest values of the averaged periodogram are at
comparatively low frequencies (the frequencies are equally spaced from 1/h
to $10^3/\text{s} = 3.6 \times 10^6\,$/h) because of slow variations in the
activity of the source.
}
\section{Periodic Bursts With Random Phase Scatter}
{The massive datasets of \citet{L21a,X21} permit consideration of the
hypothesis that there are underlying stable clocks but that bursts occur
with random phase offsets from exact periodicity.  This would be consistent
with the failure to find exact periodicity in small datasets
(Sec.~\ref{exact}).}

The differential probability distribution Prob$(\delta t)$ of an offset
$\delta t$ from an integer number of periods $P$ is taken as
\begin{equation}
	\label{cosine}
	{d\text{Prob}(\delta t) \over d\delta t} = {1 \over 2 \alpha P}
	\cos{\left({\delta t \over \alpha P}\right)}; \quad |\delta t| \le
	\pi \alpha P/2
\end{equation}
and zero otherwise.  The parameter $\alpha$ is a measure of the scatter of
the actual burst times from exact periodicity.  {This assumed functional
form is arbitrary, but has the desirable property of being an even function
of $\delta t$, peaking smoothly at $\delta t = 0$.}

The $n$-th burst occurs at a time
\begin{equation}
	\label{T}
	T_n = P N + P \alpha \sin^{-1}{(2R^\prime-1)}; \quad 1 \le n \le 
	1500,
\end{equation}
where $N = \text{NINT}(RL/P)$ is the integer closest to $R L/P$, $R$ is a
random number uniformly distributed on $[0,1)$, $P$ is taken to be $1$, $L$
(taken as $50 P$) is the total duration of the hypothesized dataset and
$R^\prime$ is another random number uniformly distributed on $[0,1)$.  The
deviations from the underlying period lie in the range $(-P\alpha\pi/2,
P\alpha\pi/2)$ with the cosine distribution Eq.~\ref{cosine}.  The full
range of scatter is $\pm \alpha \pi P/2$ and its r.m.s.~value is $\alpha
P/\sqrt{2}$.

\begin{figure}
	\centering
	\includegraphics[width=0.99\columnwidth]{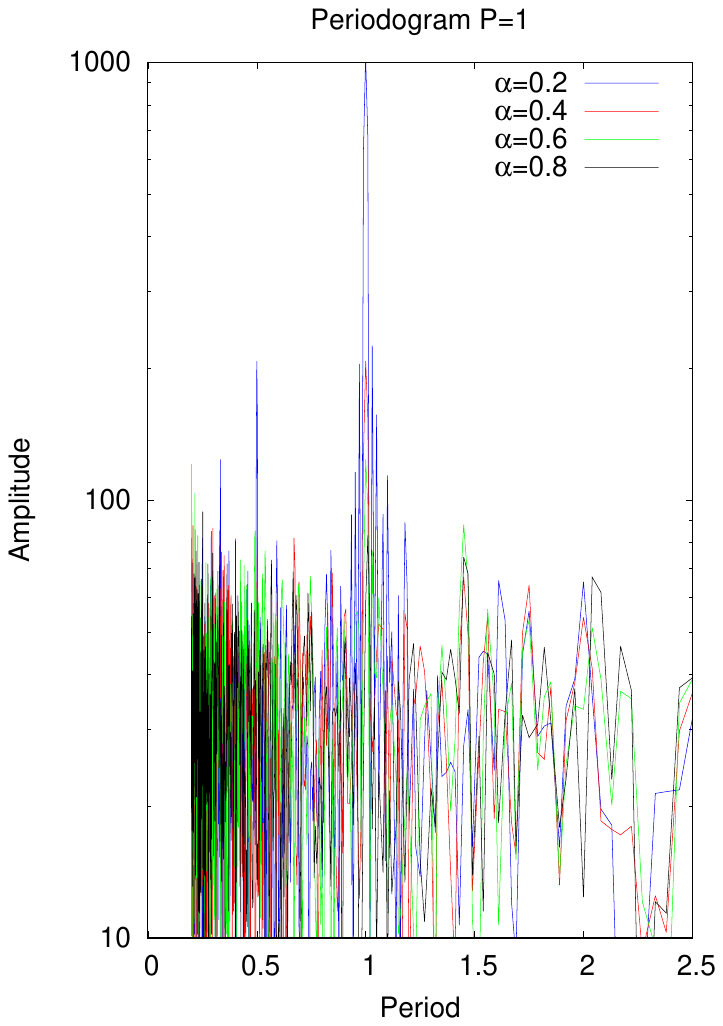}
	\caption{\label{periodograms}Periodograms for several values of
	$\alpha$.  If $\alpha \lessapprox 0.6$ there is an evident peak at
	the underlying period.}
\end{figure}

\begin{figure}
	\centering
	\includegraphics[width=2.3in]{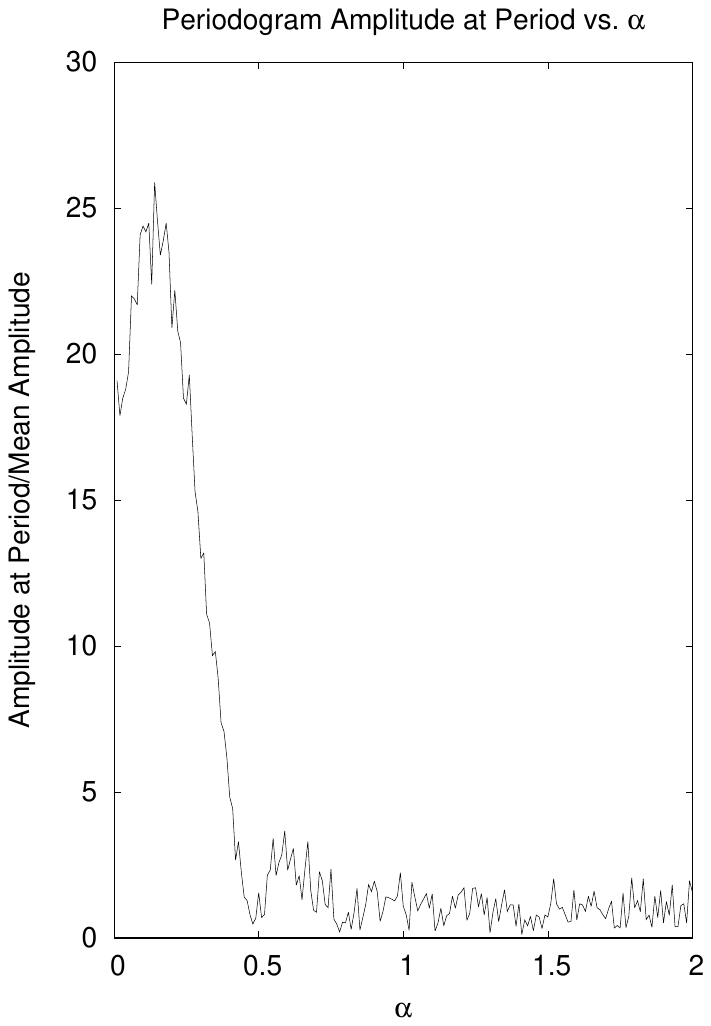}
	\caption{\label{summary} Values of the periodogram at period P,
	normalized by its mean, as a function of $\alpha$.}
\end{figure}

{The periodogram amplitude is defined by}
\begin{equation}
	A(P) = \sqrt{C^2(P)+S^2(P)},
\end{equation}
where
\begin{equation}
	\begin{split}
		C(P) &= \sum_n \cos{2 \pi T_n/P}\\
		S(P) &= \sum_n \sin{2 \pi T_n/P}.
	\end{split}
\end{equation}
The resulting periodograms are shown for several values of $\alpha$ in
Fig.~\ref{periodograms}.
Fig.~\ref{summary} shows realizations of the periodogram at the underlying
period $P$ as a function of $\alpha$, normalized to the mean periodogram for
periods from $0.01P$ to $5P$, linearly spaced in frequency.  The periodicity
is evident for $\alpha \le 0.4$.

The absence of a peak at the underlying period excludes this model for
$\alpha < 0.4$.  Somewhat larger $\alpha$ may produce a peak at that
period, although it may not be compellingly distinguished from the level of
random fluctuations.  In Fig.~\ref{periodograms} the true period is known
{\it a priori\/} while in a real dataset any true peak must be distinguished
in a statistically significant manner from random fluctuations, requiring
an amplitude greater than that shown for $\alpha = 0.6$ and possibly even
that for $\alpha = 0.4$.
\section{Bursts with Periodic Visibility Modulation}
{A rotating emitter with anisotropic emission intensity produces a peak
in the periodogram at its rotational period, because a burst is more likely
to rise above the observational threshold at certain rotational phases, and
less likely at other phases, even if the energy emitted in the bursts has no
intrinsic periodicity (as would be expected for magnetic flares).  This is
simulated using the observed burst catalogue of \citet{L21a} by discarding
a burst with probability
\begin{equation}
	\label{pdiscard}
	P_{discard} = \beta \cos^2{\left({\pi \delta t \over P}\right)},
\end{equation}
where $P$ is the period, $T_n$ is the observed epoch of the $n$-th burst and
\begin{equation}
	\delta t = T_n - P\text{NINT}(T_n/P)
\end{equation}
is the deviation of $T_n$ from exact periodicity.  Note that $|\delta t| \le
P/2$.

Discarding these bursts at assumed unfavorable phases, the maximum values of
the periodogram for all periods corresponding to evenly spaced frequencies
from 1/hour to $10^3$/s (because any period is not known {\it a priori\/} it
is necessary to consider all possible periods) are shown ini
Table~\ref{betatable}.
\begin{table}
	\begin{tabular}{|c|c|}
		\hline
		$\beta$&Peak Periodogram\\
		\hline
		0.00 & 1.706\\
		0.05 & 1.720\\
		0.10 & 1.722\\
		0.15 & 1.736\\
		0.20 & 1.798\\
		0.25 & 1.760\\
		0.30 & 1.813\\
		\hline
	\end{tabular}
	\caption{\label{betatable} Peak values of the mean over 17 observing
	sessions of the periodogram when the observed bursts of FRB 121102
	\citep{L21a} are discarded with a periodic probability, according to
	Eq.~\ref{pdiscard}.  The peak is a noisy function of $\beta$ because
	the data have temporal structure, partly random (with 1652 bursts
	randomness does not completely average out) and partly systematic
	(the activity level of FRB 121102 varies systematically with time).
	Comparison to the peak amplitude of 1.706 for the actual data (see
	also Fig.~\ref{121102mean}) shows that the probability modulation
	amplitude $\beta \lessapprox 0.20$.}
\end{table}
The large modulation that would be expected for radiation by an oblique
dipole or a more complex field distribution is not found; the data are
consistent with no periodic modulation.}
\section{Implications}
The absence of a peak in the observed periodograms, {if interpreted as a
consequence of random phase scattering,} implies $\alpha \gtrapprox 0.4$, or
a half-range of scatter $\pm 0.6 P$; the scatter from the $N$-th peak
slightly overlaps those of the $(N-1)$-th and $(N+1)$-th.  The quantitative
result depends somewhat on the assumed cosine distribution Eq.~\ref{cosine}.
{It depends chiefly on the width of the scattering rather than on the
specific functional form assumed, which must necessarily be arbitrary.
The raised cosine distribution (Hann function) used here is widely used in
statistics \citep{N81}}.   A similar result could be found for a Gaussian,
although then there would be an exponentially small (but not zero)
probability of bursts implausibly displaced by arbitrarily large amounts
from integer multiples of $P$.

The question of whether a scatter of $\pm 0.4 P$ (about $\pm 2.5$ radians)
is consistent with a rotating neutron star model depends on a detailed model
of FRB emission, which does not exist.  Emission along a bundle of field
lines emerging from a magnetic pole implies narrow collimation along the
magnetic axis (as is apparently the case in radio PSR and RRAT, as inferred
from their pulse widths), but closed field lines are typically bent by
$\sim \pi$ radians ($2\pi$ radians for the polar field lines of a dipole)
before returning to the star.  The location of the emission region is not
known, so large scatter of the emission angles is not demonstrably
impossible.  Despite this caveat, the absence of periodicity in the
observed burst arrival times suggests that aperiodic models, not based on
rotating neutron stars, should be considered.
\section{Aperiodic Models}
\label{alternatives}
Acceleration of energetic particles is nearly ubiquitous in astrophysics
\citep{K91}.  It is described phenomenologically, but there is no
fundamental understanding of why it takes place.  For example, much of
the energy output of active galactic nuclei (AGN) takes the form of
particle acceleration and the (incoherent) emission of highly relativistic
particles.  Yet it is not obvious why this is so: why does accretion onto a
supermassive black hole convert so much of the accretional power to these
nonthermal processes?  If this were not an empirical fact, we would likely
expect only the thermal emission of hot disc gas.
\subsection{Repeaters}
FRB are not emitted by AGN (although we cannot exclude the possibility of
some analogous phenomenon).  They are not emitted by stellar mass black
holes, of which there are many in our Galaxy; if they were, the Galactic
sources would dominate the FRB sky because of their proximity.  FRB sources
must be rare, and (with the possible exception of FRB 200428) no active FRB
source appears to be present in the Galaxy.  In the zoo of astronomical
objects, this suggests accreting intermediate mass black holes and their
accretion discs \citep{K17a,K19,K20}.  They are rare in the Universe, like
FRB sources, and analogy to AGN suggests the possibility of nonthermal
processes.  That analogy is also consistent with the persistent radio
sources associated with FRB 121102 \citep{M17}, FRB 190520B \citep{N21} and
FRB 20201124A \citep{R21}, and discussed more generally by \citet{L21b}.

In such a model, long-period modulation \citep{CHIME20,R20} of the activity
of a repeating FRB is readily explained as the result of precession of the
plane of the accretion disc, and hence of the direction of emission emerging
from its central funnel.  Precession is a familiar, perhaps ubiquitous,
feature of accretion discs; the original examples are Her X-1, whose disc
surrounds a neutron star \citep{K73}, and SS 433, whose disc surrounds a
stellar-mass black hole \citep{K80}.  Neither of these has been observed to
emit anything like a FRB, but their disc axes are never close to the
direction to the observer so emission close to that axis cannot be excluded.
The rarity of similar objects (there is no known analogue of SS 433 in
either our Galaxy or in any other, nor have intermediate mass black holes
been compellingly identified) is consistent with the low space density of
sources of repeating FRB, and motivates consideration of analogous systems
as their sources.

Precession of the disc axis around the orbital angular momentum axis is a
fundamental mode of oscillation of an accretion disc in a binary system, that
can be excited by irregularity or turbulence in the accretion flow, with the
companion's gravity providing the restoring force.  In addition, jitter
about the mean precession is observed \citep{KP82} and may contribute to the
aperiodicity of repeating FRB.  If FRB are emitted along the disc axis, like
the thermal plasma jets of SS 433 and the relativistic jets of AGN, the
rate at which such collimated activity will be observed depends on both the
angle between the disc axis and the direction to the observer and the
angular scatter of FRB emission about the disc axis.  Neither has been
modeled in detail, but might provide a natural explanation of the periodic
modulation of burst activity of FRB 121102 and FRB 180916 as the effect of
precession of the disc axis.
\subsection{Non-Repeaters}
Several lines of evidence indicate that non-repeating FRB differ
qualitatively from repeating FRB; they are distinguished by more than a
repetition rate.  One is the ``sad trombone'' phenomenon characteristic of
repeating FRB: within a burst the frequency of emission drifts downward,
distinguished from the effects of dispersive propagation
\citep{H19,J19,Ra20}.  Its origin is not understood, but it may be a useful
phenomenological tool for distinguishing the two classes of sources even if
only one burst has been observed.  Other distinguishing characteristics have
been reported by \citet{A21,P21}.

Another argument is the bimodality of the duty factors, defined by
\begin{equation}
	D \equiv {\langle F \rangle^2 \over \langle F^2 \rangle},
\end{equation}
where $F$ is the flux.  For repeaters $D$ is typically $\sim 10^{-5}$ while
for non-repeaters (where only an upper limit can be found) $D \lesssim
10^{-8}\text{--}10^{-10}$ for the best observed non-repeating FRB
\citep{K17b,K18,K19}.  If repeating and non-repeating FRB differed only
quantitatively, a monotonic, rather than bimodal, distribution of $D$ would
be expected.

Arguments derived from the periodograms of repeating FRB are inapplicable to
non-repeaters.  But arguments based on the space density of FRB sources do
apply to non-repeaters, and have been considered by \citet{H20}.  The
volumetric rate of observed non-repeating FRB in the local ($z \lesssim 1$)
universe approaches or exceeds that of known or plausibly inferred classes
of catastrophic events, such as stellar collapses or mergers.  The observed
volumetric FRB rate is surely an underestimate of its true value, that is
much larger if their emission is beamed, if there is a population of FRB
below instrumental detection thresholds.  It is {\it a priori\/} implausible
that the distribution of burst fluxes or fluences cuts off just below
instrumental sensitivities (this would be inconsistent with the observation
that both cosmologically distant and cosmologically local FRB have fluxes
and fluences ranging down to detection thresholds; they are not standard
candles), or if FRB distances are overestimated because of
non-intergalactic contributions to their dispersion measures \citep{N21}.
The observed rate of apparently non-repeating FRB can only set a lower bound
on the rate of catastrophic events to which they are attributed because it
is not determinable how many FRB-like events fall below the detection
threshold.  Almost any downward extrapolation of the flux or fluence
distribution would indicate that the ``true'' rate must be much higher than
the observed rate.  

Unless there is a completely unanticipated class of catastrophic events,
appeal must be made to less catastrophic events that might, in principle,
repeat, but at a much lower rate than those of known repeating FRB.  The
giant outbursts of SGR are not extrapolations of their smaller eruptions,
but are qualitatively different outliers \citep{K21a}.  This is attributed to
a global rearrangement of the magnetic geometry, analogous to a crack that
propagates through the entire neutron star crust, while the smaller
eruptions are attributed to localized reconnection and flares.  The giant
outburst of SGR 1806$-$20 did not make a FRB \citep{TKP16}, likely because
it filled the magnetosphere with opaque and electrically conducting
equilibrium pair plasma \citep{K96}.  Beaming is an unsatisfactory
explanation because it would require 11 orders of magnitude difference
between in-beam and out-of-beam emission, inconsistent with even a tiny
amount of scattering or extension of the emission region along a field line.

A less energetic but global rearrangement, perhaps of a magnetosphere with
lower field, might similarly be a statistical outlier (not the high energy
tail of a smooth distribution of smaller eruptions) without creating a dense
pair plasma that would suppress FRB emission.  It is impossible to estimate
the recurrence time of such bursts (the recurrence time of SGR is at least
a few decades), but it might well be $> 1\,$y, making it difficult to
observe repetitions, but $\ll 10^{10}\,$y, permitting an event rate orders
of magnitude greater than the birth rate of FRB sources and the rates of
stellar collapse, merger or similar catastrophic events \citep{H20}.  The
development of instruments capable of large angular acceptance angle or
all-sky monitoring of FRB will offer the possibility of observing
repetition rates $< 1\,$/y.
\section{Discussion}
The failure to detect periodicity in repeating FRB casts doubt on popular
models that attribute them to rotating magnetic neutron stars.  The data
exclude not only strict, PSR-like or RRAT-like periodicity, but even models
with substantial scatter of burst times around rotations of a presumed
neutron star source.  Aperiodic models, such as accretion discs around black
holes should be considered, even though there is no understood mechanism by
which they might produce FRB (a problem also with neutron star models).

Black hole accretion disc models must address the absence of FRB from known
black holes, both stellar mass black holes in binaries and supermassive
holes in AGN.  This might be related to the mass of the black hole, which is
why intermediate mass black holes are suggested, although no mechanism for
this is evident.

There may be large observational selection effects.  The inner regions of
the accretion discs and funnels of luminous stellar mass and AGN black holes
are in the intense radiation fields that make these objects observable; such
radiation fields are hostile environments for the acceleration of energetic
electrons because of Compton scattering.  If only a small fraction of black
hole accretion discs are favorably oriented for observation of FRB, and FRB
are preferentially (or only) emitted by low luminosity black hole accretion
then observational selection mitigates against the identification of FRB
sources with identified black holes.

The related environments of FRB 121102 and SGR/PSR J1745-2900 \citep{K21b}
suggests this latter object as a candidate repeating FRB.  No FRB has been
observed from it, but its Galactic location means that even micro-FRB would
be detectable.  It might repay monitoring.

Non-repeating FRB pose a different problem because the rate of
catastrophic events is insufficient, even if each catastrophe produces an
observable FRB.  This suggests appealing to intrinsically infrequent but
repeating events that are not accompanied by a larger number of weaker
events, such as would be implied by a power law distribution of their
fluxes or fluences, in analogy to the giant outbursts of SGR.
\section*{Acknowledgment}
I thank Alex Chen, Casey Law and Tsvi Piran for useful discussions, {Wang
Pei for providing data in computer-readable format, Lilly M. Canel-Katz for
assistance in data processing and an anonymous referee for pointing out the
importance of phase drift in periodograms of long time series}.
\section*{Data Availability}
This theoretical study did not generate any new data.  Codes and their
output will be provided upon request.

\appendix
\section{Highest amplitudes in periodograms}
{
\begin{table}
	\centering
	\begin{tabular}{|rrrr|}
		\hline
		Session & MJD & Duration & \# Bursts \\
		\hline
		1 & 58724 & 3 h & 87 \\
		2 & 58725 & 3 h & 121 \\
		3 & 58726 & 4 h & 110 \\
		4 & 58727 & 5 h & 91 \\
		5 & 58728 & 3 h & 65 \\
		6 & 58730 & 1 h & 122 \\
		7 & 58733 & 1 h & 81 \\
		8 & 58738 & 1 h & 58 \\
		9 & 58746 & 1 h & 52 \\
		10 & 58748 & 1 h & 53 \\
		11 & 58749 & 1 h & 50 \\
		12 & 58752 & 1 h & 54 \\
		13 & 58753 & 1 h & 53 \\
		14 & 58754 & 1 h & 60 \\
		15 & 58756 & 1 h & 117 \\
		16 & 58757 & 1 h & 64 \\
		17 & 58758 & 1 h & 53 \\
		\hline
\end{tabular}
	\caption{\label{sessions} Observing sessions from \citet{L21a} used
	in the analysis.  MJD are truncated to whole days and refer to the
	beginning of the session; some extend into the next MJD.  Durations
	are rounded to the closest whole hour.}
\end{table}

\begin{table*}
	\centering
	\begin{tabular}{|rc|rc|rc|rc|rc|rc|}
		\hline
		$\nu$ (/h) & $A$ & $\nu$ (/h) & $A$ &$\nu$ (/h) & $A$ & $\nu$ (/h) & $A$ &$\nu$ (/h) & $A$ & $\nu$ (/h) & $A$ \\
		\hline
  47901&  1.706& 148579&  3.855&  91179&  4.102& 378426&  4.181& 175612&  4.059&  11196&  4.053\\
  17329&  1.656&1071898&  3.758&  69896&  4.095& 162156&  4.023& 549345&  3.992&1599906&  3.999\\
   9624&  1.622&2682736&  3.738& 170499&  3.867&   3148&  4.007&   6499&  3.915&  40569&  3.693\\
   5389&  1.609&  48531&  3.661& 141895&  3.809& 248731&  3.966& 129301&  3.758& 616044&  3.635\\
  26551&  1.609&3198871&  3.634&2614574&  3.766&  64707&  3.957&3450210&  3.690&  29379&  3.605\\
   4445&  1.608& 931421&  3.616&2837357&  3.733& 142695&  3.953&    121&  3.667&  15347&  3.598\\
  21496&  1.605& 299359&  3.598& 123156&  3.716& 178039&  3.934&3545877&  3.623&  15659&  3.583\\
   2807&  1.601&2279492&  3.591&  11118&  3.702& 105138&  3.933&   8576&  3.609& 625709&  3.573\\
  77126&  1.582&3541614&  3.528& 601611&  3.695& 195010&  3.846&3050026&  3.605&3204853&  3.557\\
  61291&  1.565&1437774&  3.518&3040658&  3.683&3575491&  3.787&  51698&  3.595&2329041&  3.506\\
  \hline
 159675&  4.244& 189010&  4.466&3569504&  3.686& 548541&  4.133&  19787&  3.824&   1511&  4.066\\
3507651&  3.840&   1788&  4.275&  22776&  3.653&2668467&  3.950&1808884&  3.671&  50929&  3.799\\
 375944&  3.817&   7436&  4.173&   8272&  3.639& 141965&  3.762&  30795&  3.668&  31241&  3.740\\
2556501&  3.713&  16137&  4.128&1097262&  3.594&2237014&  3.646& 779916&  3.614&  93852&  3.737\\
  32369&  3.690&  73887&  4.064&  13328&  3.581& 731623&  3.646& 200291&  3.593&  38728&  3.661\\
3370705&  3.664& 770860&  3.962&2070272&  3.580&2004880&  3.593&2940895&  3.557&   8160&  3.608\\
1096479&  3.623& 431879&  3.932&2515606&  3.579& 702377&  3.576&1333734&  3.543& 101220&  3.591\\
 178908&  3.608&  10203&  3.928& 996211&  3.560&2822125&  3.573&  45791&  3.536& 104497&  3.576\\
3467095&  3.602& 276665&  3.924&2449052&  3.504&  24339&  3.570&  78545&  3.528&3310589&  3.575\\
  62824&  3.600&3069486&  3.905&1797786&  3.476&2950469&  3.567& 110489&  3.508& 787970&  3.572\\
  \hline
2795706&  3.920& 146837&  3.930&1078954&  3.783& 867699&  4.022&   1046&  4.021&   5688&  3.834\\
2657514&  3.905&  34684&  3.707&  47087&  3.682&   2509&  3.944&1587639&  3.922&3161165&  3.671\\
2829638&  3.658&2741838&  3.616&1057309&  3.610&   5733&  3.878& 828385&  3.734&3316185&  3.666\\
2295030&  3.653&2762144&  3.589&2494837&  3.589&  40230&  3.816&  27757&  3.714& 102691&  3.656\\
2770545&  3.640&1489132&  3.572&2498414&  3.544&1703863&  3.777&2361002&  3.710& 673147&  3.563\\
1566664&  3.598&1473871&  3.560&3539423&  3.529&  52659&  3.769& 806658&  3.684& 163929&  3.546\\
   4820&  3.595&2531777&  3.558& 104813&  3.500&  98413&  3.763&3042096&  3.642&   8669&  3.538\\
2596561&  3.586&  71442&  3.549&   8233&  3.496&  92342&  3.755&    266&  3.623&  63167&  3.533\\
2469484&  3.519&1404559&  3.548&1064584&  3.474& 115360&  3.746&  22477&  3.610& 960611&  3.532\\
   3850&  3.513&2527031&  3.544&2028999&  3.469&  48365&  3.726&  65864&  3.601&  61964&  3.528\\
		\hline
\end{tabular}
	\caption{\label{Amptable} Amplitudes $A$ (arbitrary units) of
	largest ten elements of the average (top row, left) and 17
	individual session periodograms, with periods from 1 ms to 1 h, of
	FRB 121102 bursts, from data of \citet{L21a}.  The extreme values of
	the averaged periodogram are less than those of the individual
	session periodograms because at a frequency at which one periodogram
	has an extreme value, others generally will not.}
\end{table*}
}
\label{lastpage}
\end{document}